\documentstyle[prb,aps,epsfig,floats,eqsecnum,amsmath,amssymb]{revtex}

\begin{document}
\draft
\twocolumn[\hsize\textwidth\columnwidth\hsize\csname@twocolumnfalse\endcsname
 
\title{Magnetic order in the Shastry-Sutherland model}
\author{A. Fledderjohann, K.-H. M{\"u}tter}
\address{Physics Department, University of Wuppertal, 42097 Wuppertal, Germany}
\maketitle                                                                         
%
\begin{abstract}
The ground state properties of the Shastry-Sutherland model in the
presence of an external field are investigated by means of variational
states built up from unpaired spins (monomers) and singlet pairs of
spins (dimers). The minimum of the energy is characterized by specific
monomer-dimer configurations, which visualize the magnetic order in
the sectors with fixed magnetization $M=S/N$. A change in the
magnetic order is observed if the frustrating coupling $\alpha$
exceeds a critical value $\alpha_c(M)$, which depends on $M$.
Special attention is paid to the ground state configurations at
$M=1/4, 1/6, 1/8$. 
\end{abstract}
%
\pacs{75.10.-b, 75.10.Jm}
\twocolumn]
 
\section{Introduction}
The Shastry-Sutherland model\cite{shastry81} defined by the two-dimensional 
spin 1/2 Hamiltonian
 
\begin{eqnarray}
\label{h0}
H & = & \sum_{\langle {\bf x},{\bf y}\rangle} {\bf S}({\bf x})
{\bf S}({\bf y})+\alpha
\sum_{\langle\langle {\bf x},{\bf y}\rangle\rangle}{\bf S}({\bf x})
{\bf S}({\bf y})
\end{eqnarray}
 
with nearest neighbor couplings and frustrating next-nearest neighbor
couplings on the diagonals shown in Fig. \ref{fig1}  has attracted a
lot of interest for theoretical and experimental reasons:
 
\begin{figure}[ht!]
\centerline{\epsfig{file=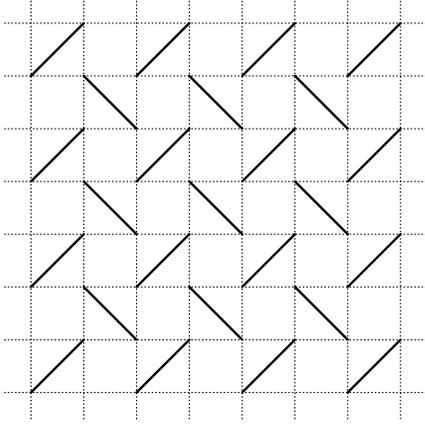,width=7.0cm,angle=0}}
\caption{The couplings in the Shastry-Sutherland model. Nearest
and next-nearest neighbor couplings are represented by dotted
and solid lines.}
\label{fig1}
\end{figure}
 
(1) The product wave function
 
\begin{eqnarray}
\label{p_wf1}
\Psi & = & \prod_{\langle\langle {\bf x},{\bf y}\rangle\rangle}
[{\bf x},{\bf y}]
\end{eqnarray}
 
built up from singlet states
 
\begin{eqnarray}
\label{p_wf}
[{\bf x},{\bf y}] & = & \frac{1}{\sqrt{2}}\left(\chi_+({\bf x})
\chi_-({\bf y})-\chi_-({\bf x})\chi_+({\bf y})\right)
\end{eqnarray}
 
is known to be an eigenstate of the Hamiltonian \eqref{h0}, which
turns out to be the ground state, if the coupling $\alpha$
exceeds a critical value $\alpha_c$ ($\alpha_c\approx 1.4$).
\cite{miyahara99} The phase diagram has been studied recently
by Weihong, Oitmaa and Hamer \cite{weihong01} by means of
series expansions.
 
(2) The Hamiltonian \eqref{h0} is suggested to be an appropriate
model for the compound $SrCu_2(BO_3)_2$ the magnetic properties
of which have been investigated in recent high magnetic field
experiments.\cite{kageyama99,onizuka00,nojiri99}
Plateaus have been found in the magnetization
curve $M=M(B)$ at rational values of the magnetization
$M/M_S=1/3,\,1/4,\,1/8$, where $M_S=1/2$ is the saturating
magnetization.
 
From the theoretical point of view the appearance of magnetization
plateaus is well understood in quasi one-dimensional systems, e.g.
with ladder geometry. \cite{wiessner00}
Here, a quantization rule has been formulated 
by Oshikawa, Yamanaka and Affleck, \cite{oshikawa97a} which originates
from the prediction of soft modes \cite{fl99} based on the 
Lieb-Schultz-Mattis (LSM) theorem. \cite{lieb61}
Only the position of the possible plateaus -- i.e. the quantized
value of the magnetization -- are predicted by this rule. 
The upper and lower critical fields which define the width of
the plateau, however, depend on the magnitude of transition
matrix elements with a momentum transfer corresponding to the
relevant soft mode. These matrix elements contribute to the
dynamical and static structure factors and a strong peak in
these quantities at the soft mode momenta is needed for a
pronounced plateau in the magnetization curve. \cite{fl99}

The extension of the Lieb Schultz Mattis construction to higher
dimensions ($D>1$) meets difficulties. As was pointed out by
Oshikawa \cite{oshikawa00} magnetization plateaus are possible
in higher dimensions as well, provided that the ``commensurability
condition'' is satisfied. Based on a topological argument he shows,
that this condition is a robust non-perturbative constraint.



The emergence of magnetization plateaus in a modified
Shastry-Sutherland model has been discussed in
Ref. [\onlinecite{mueller-hartmann00}]. Recently, Misguich, Jolic\oe r,
and Girvin studied the emergence of magnetization plateaus in the
framework of a Chern-Simons theory.\cite{misguich01}

In Ref. [\onlinecite{kageyama99}]
Kageyama et al. proposed that product wave functions of the type 
\eqref{p_wf1} with certain distributions of $N_S$ singlets \eqref{p_wf}
and $N_T$ triplets

\begin{eqnarray}
\{{\bf x},{\bf y}\} & = & \chi_+({\bf x})\chi_+({\bf y})
\end{eqnarray}

might yield an appropriate ansatz for the ground state in the sector
with total spin

\begin{eqnarray}
S_T & = & N_T/4
\end{eqnarray}

where $N_T$ and $N_S$ are constrained by the total number of sites

\begin{eqnarray}
N_S+N_T & = & N/2\,.
\end{eqnarray}
Typical examples of these states are shown in Figs. \ref{fig2}(b),
\ref{fig3}(b), \ref{fig4}(c) and \ref{fig4}(d).
An effective Hamiltonian, describing the interaction between
singlets and triplets, has been developed and evaluated in Refs.
[\onlinecite{momoi00,fukumoto00}].

In this paper, we investigate a wider class of product wave functions
- which we call monomer-dimer configurations \cite{karbach93} - and 
which are aimed to
describe the magnetic order at those magnetizations ($M=1/4,1/6,1/8,1/16$),
where plateaus are expected. Indeed we find a change in the magnetic
order if the frustration parameter exceeds a critical value 
$\alpha_c(M)$, which depends on the magnetization $M$. For
$\alpha>\alpha_c(M)$ we recover the singlet-triplet configurations
proposed in Ref.[\onlinecite{kageyama99}]. 
For $\alpha<\alpha_c(M)$,
however, we find new configurations with lower energy.

The outline of the paper is as follows: In Sec. II we define the
monomer-dimer configurations. In Sec. III we minimize the
expectation value of the Hamiltonian \eqref{h0} between 
monomer-dimer configurations. This procedure singles out specific
configurations, which visualize the magnetic order at fixed
magnetization $M$. 
In Sec. \ref{frozen_monomer} we introduce the ``frozen monomer
approximation'', which allows to lower the energy expectation
values between monomer-dimer configurations without changing the
magnetic order, i.e. the distribution of ``frozen'' monomers.

The quality of the frozen monomer approximation is studied for 
$M=1/8,1/6,1/4$ in Sec. \ref{numerics} by a comparison with the
ground state energies obtained from exact diagonalizations on
finite clusters.

We also look for the formation of magnetization plateaus.
Possible interpretations of the observed plateaus in
$SrCu_2(BO_3)_2$ are discussed in Sec. VI.

 
\section{Monomer-Dimer configurations}
 
The Shastry-Sutherland Hamiltonian \eqref{h0} conserves the total spin
 
\begin{eqnarray}
{\bf S} & = & \sum_{\bf x}{\bf S}({\bf x})
\end{eqnarray}
 
and we therefore start from eigenstates of

\begin{eqnarray}
{\bf S}^2 & = & S(S+1) \quad\mbox{and}\quad S_3\,=\,-S,\ldots,S\,.
\end{eqnarray}                                                                                

Following Hulth´en \cite{hulthen38}, these states $|K,\nu=2S\rangle$
can be constructed in the sector with total spin $S$ -i.e. magnetization
$M=S/N$- as product states of

\begin{itemize}
\item
unpaired spin-up states at sites ${\bf x}_1,\ldots,{\bf x}_{\nu}$,
$\nu=2S$:
\begin{eqnarray}
\label{monomer}
|{\bf x}+\rangle & = & \chi_+({\bf x})
\end{eqnarray}
which we call ``monomers''
\item
singlets of paired spins $[{\bf x},{\bf y}]$ \eqref{p_wf} at sites
${\bf x}$, ${\bf y}$ (``dimers''):
\begin{eqnarray}
\label{product_ansatz}
|K,\nu\rangle & = & \prod_{j=1}^{\nu=2S}|{\bf x}_j+\rangle
\prod_{\langle{\bf x},{\bf y}
\rangle} [{\bf x},{\bf y}]
\end{eqnarray}  
\end{itemize}

Note, that in the monomer-dimer configuration $K$ each site ${\bf x}$ is
occupied exactly once: either by a monomer or a dimer.
Moreover, monomer-dimer configurations $|K,\nu\rangle$ yield an overcomplete
non-orthogonal set of eigenstates with total spin $S$. 

The expectation value of the Hamiltonian \eqref{h0} between monomer-dimer
configurations can be easily calculated with the following rules:

\begin{eqnarray}
\langle 1+,2+|{\bf S}(1){\bf S}(2)|1+,2+\rangle & = & \frac{1}{4}\\
\langle [1,2]|{\bf S}(1){\bf S}(2)|[1,2]\rangle & = & -\frac{3}{4}\\[2pt]
\label{single_monomer}
\langle [1,3]2+|{\bf S}(1){\bf S}(2)|[1,3]2+\rangle & = & 0\\[6pt]
\langle [1,3][2,4]|{\bf S}(1){\bf S}(2)|[1,3][2,4]\rangle & = & 0\,.
\end{eqnarray}  

If we count on each configuration the numbers

\begin{itemize}
\item    
$N_1^{(0)}(K)$ of nearest neighbor dimers
\item    
$N_2^{(0)}(K)$ of next-nearest neighbor dimers (corresponding to 
Fig. \ref{fig1})
\item
$N_1^{(1)}(K)$ of nearest neighbor monomer pairs
\item
$N_2^{(1)}(K)$ of next-nearest neighbor monomer pairs (corresponding 
to Fig. \ref{fig1})
\end{itemize} 
we can immediately compute the expectation value:

\begin{eqnarray}
\label{khk}
\langle K,\nu |H|K,\nu\rangle & = & -\frac{3}{4}N_1^{(0)}-
\frac{3}{4}\alpha N_2^{(0)}\nonumber\\
 & & +\frac{1}{4}N_1^{(1)}+\frac{\alpha}{4}N_2^{(1)}\,.\label{expec}
\end{eqnarray}

In order to minimize this expectation value we have to look for 
configurations with
\begin{itemize}
\item          
a maximum number of nearest neighbor dimers $N_1^{(0)}$ if $\alpha<1$
\item
a maximum number of next-nearest neighbor dimers $N_2^{(0)}$ if $\alpha>1$
\item
a minimum number of monomer pairs $N_1^{(1)}$, $N_2^{(1)}$ on nearest
and next-nearest neighbor sites.
\end{itemize}

 
\section{Magnetic ordering at fixed magnetizations}

\subsection{$M=1/4$, $\nu=N/2$}
 
\begin{figure}[ht!]
\centerline{\epsfig{file=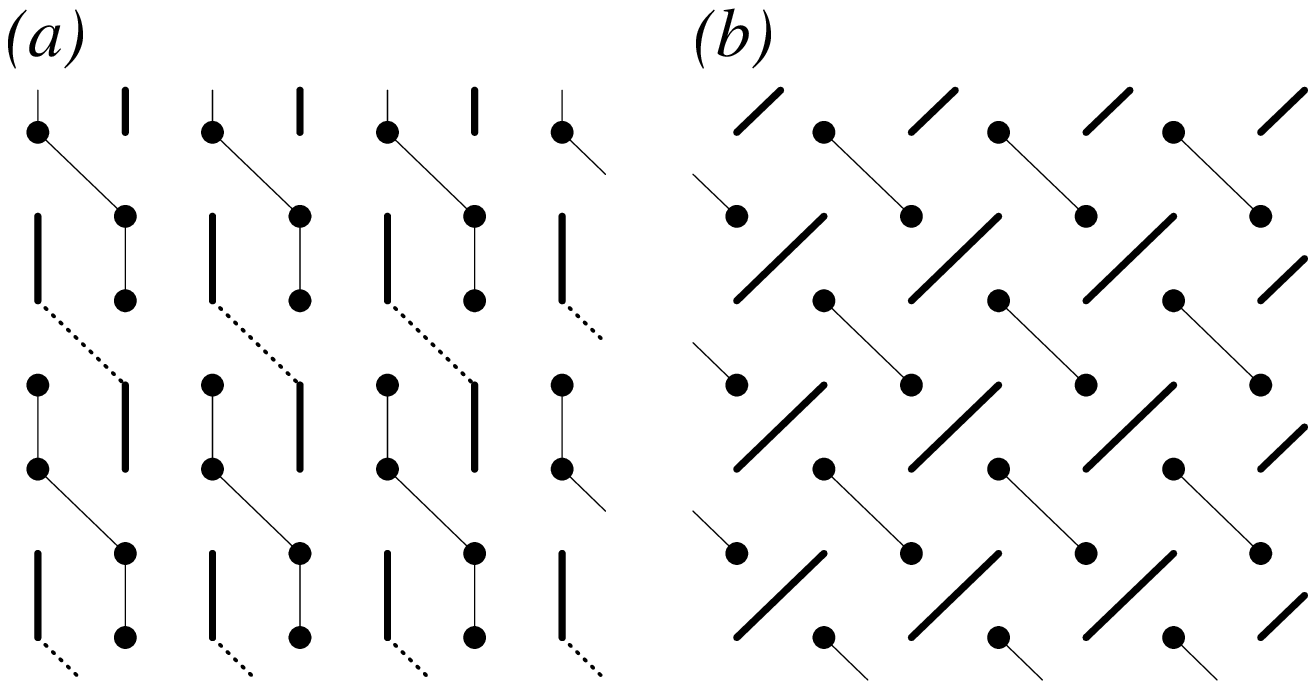,width=8.5cm,angle=0}}
\caption{Monomer-dimer configurations with minimal energy expectation
value at $M=1/4$. Bold lines represent ``dimers'' i.e. paired spins
coupled to singlets. Dotted lines indicate the couplings between
dimers induced by Hamiltonian \eqref{h0}. Unpaired spin-up states
(``monomers'') [Eq.\eqref{monomer}] are symbolized by solid points;
their couplings on nearest and next-nearest neighbor sites are
indicated by thin lines. (a) configuration $K_1$ for 
$\alpha<\alpha_c(M=1/4)=4/5$; (b) configuration $K_2$ for
$\alpha>\alpha_c(M=1/4)$}
\label{fig2}
\end{figure}

Let us start with $M=1/4$. In this situation we have to distribute
$\nu=N/2$ monomers and $N/4$ dimers on the square lattice. We cannot
avoid the appearance of monomer pairs on nearest and next-nearest
neighbor sites, but we can minimize their numbers $N_1^{(1)}$ and
$N_2^{(1)}$ if we cover the lattice in the way shown in 
Fig.\ref{fig2}(a). The bold lines  symbolize the nearest-neighbor
singlets, the thin lines the monomer pairs on nearest and next-nearest
neighbor sites. Dimer pairs can interact via the next-nearest
neighbor couplings in the Hamiltonian; they are represented by
dotted lines. According to \eqref{expec}, the expectation value of
the Hamiltonian is found to be:

\begin{eqnarray}
\label{e1m1_4}
E(K_1,1/4) & = & \langle K_1,N/2 |H|K_1,N/2\rangle\nonumber\\
 & = & \frac{N}{8}(-1+\alpha/4)\,.
\end{eqnarray}  

A second configuration, shown in Fig. \ref{fig2}(b) has been proposed
in Ref.[\onlinecite{onizuka00}]
as a possible ground state configuration for $M=1/4$. In this case,
the expectation value of $H$ turns out to be:

\begin{eqnarray}
\label{e2m1_4}
E(K_2,1/4) & = & \langle K_2,N/2 |H|K_2,N/2\rangle\nonumber\\
 & = & -\alpha\frac{N}{8}\,.
\end{eqnarray}  

The difference of \eqref{e1m1_4} and \eqref{e2m1_4}
\begin{eqnarray}
E(K_1,1/4)-E(K_2,1/4) & = & \frac{N}{8}(-1+\frac{5}{4}\alpha)\nonumber\\
 & & 
\end{eqnarray}
changes its sign for $\alpha=4/5$, which means there is a change in
the magnetic order from configuration $K_1$ to $K_2$ if the 
frustration parameter exceeds the value $\alpha_c(M=1/4)=4/5$.

\subsection{$M=1/6$, $\nu=N/3$}

\begin{figure}[ht!]
\centerline{\epsfig{file=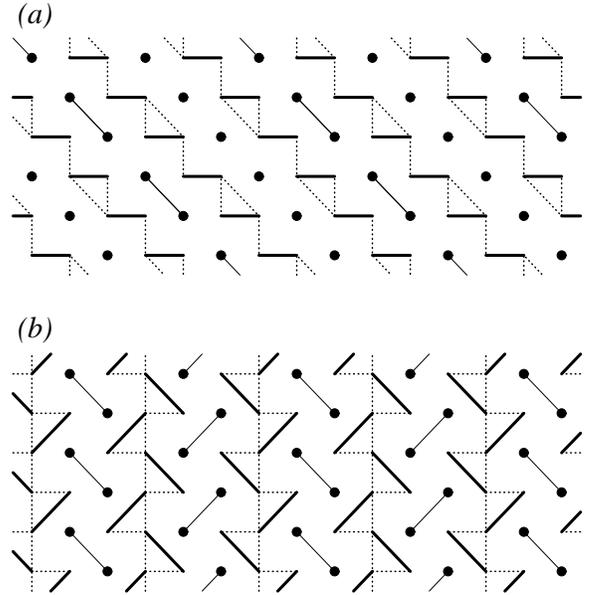,width=8.5cm,angle=0}}
\caption{Same as Fig.\ref{fig2} for $M=1/6$: (a) configuration
$K_1$ $\alpha<\alpha_c(M=1/6)=12/11$; (b) configuration $K_2$
for $\alpha>\alpha_c(M=1/6)$}
\label{fig3}
\end{figure}

Next, we turn to the case $M=1/6$, where we have to distribute $\nu=N/3$
monomers and $N/3$ dimers on the lattice. The configuration $K_1$
[Fig. \ref{fig3}(a)]
minimizes the number $N_2^{(1)}$ of monomer pairs on next-nearest
neighbor sites, whereas in the configuration $K_2$
[Fig. \ref{fig3}(b)] the next-nearest
neighbor sites of Fig. \ref{fig1} are occupied with singlets and
triplets in the spirit of Ref.[\onlinecite{onizuka00}]. The 
difference of the expectation values of $H$

\begin{eqnarray}
E(K_1,1/6) & = & \langle K_1,N/3 |H|K_1,N/3\rangle\nonumber\\
 & = & -\frac{N}{4}(1-\frac{\alpha}{12})\\
E(K_2,1/6) & = & \langle K_2,N/3 |H|K_2,N/3\rangle\nonumber\\
 & = & -\frac{N}{4}\frac{5\alpha}{6}
\end{eqnarray}
\begin{eqnarray}
E(K_1,1/6)-E(K_2,1/6) & = & -\frac{N}{4}(1-\frac{11}{12}\alpha)\nonumber\\
 & & 
\end{eqnarray}
changes sign for $\alpha_c(M=1/6)=12/11$. Again we observe a change in the
magnetic order from $K_1$ to $K_2$ if $\alpha$ passes this value.

It is remarkable to note, that in both cases $\alpha<\alpha_c(M=1/6)$
and $\alpha>\alpha_c(M=1/6)$ a stripe order of the monomers is
predicted. 


\subsection{$M=1/8$, $\nu=N/4$}

In the case of $M=1/8$ we have to distribute $\nu=N/4$ monomers and
$3N/8$ dimers on the square lattice. We can avoid now completely
the appearance of monomer pairs on nearest and next-nearest
neighbor sites as is demonstrated by the configuration $K_1$
shown in Fig. \ref{fig4}(a). Owing to the stripe structure,
we can also construct a second configuration [Fig. \ref{fig4}(b)]
with $N_1^{(0)}=N/8$ dimers on nearest neighbor sites and $N_2^{(0)}=N/4$
dimers on next-nearest neighbor sites. Two further configurations $K_3$ and 
$K_4$ have been proposed in Refs. [\onlinecite{onizuka00,fukumoto00}], 
which only contain $N_1^{(0)}=3N/8$
dimers and $N_1^{(1)}=N/8$ monomer pairs on next-nearest neighbor sites.


The corresponding energy expectation values are
\begin{eqnarray}
E(K_1,1/8) & = & -\frac{9}{32}N\label{e1m1_8}\\
E(K_2,1/8) & = & -\frac{3}{32}N(1+2\alpha)\\
E(K_3,1/8) & = & E(K_4,1/8)=-\frac{N}{4}\alpha
\label{e3m1_8}\,.
\end{eqnarray}
Comparing the expectation values \eqref{e1m1_8}-\eqref{e3m1_8} we
expect a change in the magnetic order with $\alpha$:

\begin{figure}[ht!]
\centerline{\epsfig{file=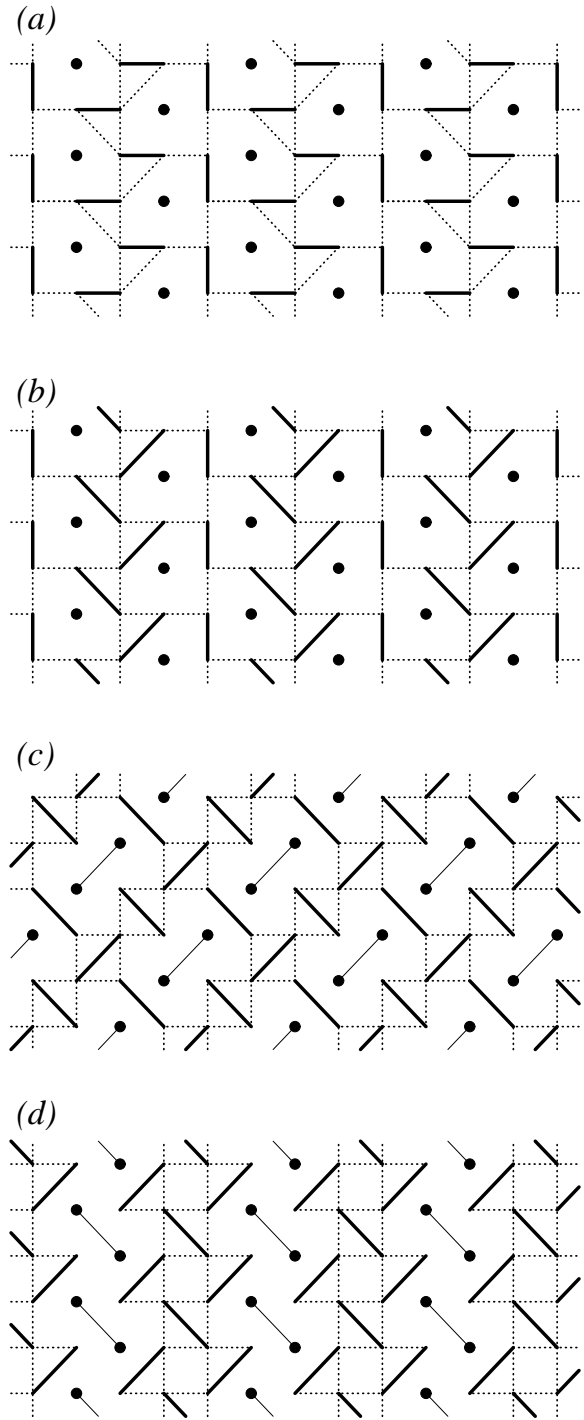,width=8.5cm,angle=0}}
\caption{Same as Fig.\ref{fig2} for $M=1/8$: (a) configuration
$K_1$, $\alpha<1$; (b) configuration $K_2$, $1<\alpha<3/2$; (c)
and (d) configurations $K_3$ and $K_4$, $\alpha>3/2$}
\label{fig4}
\end{figure}

\begin{eqnarray}
\alpha<1:\quad & & E(K_1)<E(K_2)<E(K_3)=E(K_4)\\
1<\alpha<\frac{3}{2}: & & E(K_2)<E(K_1)<E(K_3)=E(K_4)\\
\frac{3}{2}<\alpha:\quad & & E(K_3)=E(K_4)<E(K_2)<E(K_1)\,.
\end{eqnarray}


\section{The frozen monomer approximation (FMA)}
\label{frozen_monomer}

The monomer-dimer configurations which we developed in the last
section to describe the magnetic order in the Shastry-Sutherland
model are not eigenstates of the Hamiltonian \eqref{h0}. Application
of \eqref{h0} onto these states will generate new states. In this
section we study the impact of those couplings in the Hamiltonian,
which generate interactions only between dimer pairs, i.e. we
consider an approximation where the $\nu$ monomers are frozen at
sites ${\bf x}_1\ldots{\bf x}_{\nu}$ in the configuration
$|K,\nu\rangle$. For each of these configurations, we define a 
decomposition of the Hamiltonian in three parts:

\begin{eqnarray}\label{a1}
H & = & H_{\nu}+ H(K)+\sum_{i}H({\bf x}_i,K)
\end{eqnarray}
where
 
\begin{itemize}
\item[a)]
$H_{\nu}$ contains all the nearest and next-nearest neighbor
couplings between the sites ${\bf x}_1\ldots{\bf x}_{\nu}$,
where the monomers are located. All other sites are occupied
with dimers. They form an antiferromagnetic cluster $K$, which
are represented by the dimers and the dotted connections 
between dimers in Figs. \ref{fig2}(a)-\ref{fig4}(d).
 
\item[b)]
The cluster Hamiltonians $H(K)$ is defined by the nearest
and next-nearest neighbor couplings on the dimer cluster $K$.
 
\item[c)]
The nearest and next-nearest neighbor couplings in $H({\bf x}_i,
K)$ take into account the remaining interactions between
the monomer at site ${\bf x}_i$ and the dimers in the cluster $K$.
 
\end{itemize}

The ground state energy $E(K)$ of the antiferromagnetic cluster
Hamiltonian $H(K)$
\begin{eqnarray}\label{a4}
H(K)\Psi(K) & = & E(K)\Psi(K)\,.
\end{eqnarray}
is obviously lower than the expectation value of $H(K)$ between
the dimer product wave function on the cluster $K$.

The product ansatz
\begin{eqnarray}
\label{a6}
|{\bf x}_1\ldots{\bf x}_{\nu},K\rangle & = &
\prod_{i=1}^{\nu=2S}|{\bf x}_i+\rangle\prod_j\Psi(K)
\end{eqnarray}
yields an eigenfunction of $H_{\nu}+H(K)$ with energy
\begin{eqnarray}
\label{a8}
E(K,\nu) & = & \frac{1}{4}N_1^{(1)}(K)+\frac{\alpha}{4}
N_2^{(1)}(K)+E(K)\,
\end{eqnarray}
which again represents an upper bound for the exact ground
state energy $E_0(M=\nu/2N)$ of \eqref{a1} in the sector
with magnetization $M=S/N$:
\begin{eqnarray}
\label{a10}
E_0(M=\nu/2N) & \leq & E(K,\nu)\,.
\end{eqnarray}
In the derivation of \eqref{a10} one has to use the fact,
that the expectation value of the interaction term $H(\nu,K)$
between the product state \eqref{a6} vanishes, since
\begin{eqnarray}
\label{a11}
\langle\Psi(K)|S_l(y)|\Psi(K)\rangle & = & 0\quad y\in K,\,\, l=1,2,3\,.
\end{eqnarray}

\section{Numerical results}
\label{numerics}

In order to check the quality of the frozen monomer approximation (FMA),
we have computed the energies \eqref{a8} and compared with exact
diagonalizations of the Shastry-Sutherland Hamiltonian at fixed
magnetization $M=\nu/2N$ on lattices
with $N=4\times 4=16$ and $N=4\times 6=24$.

The strongest effects due to the frozen monomer approximation occur
at small magnetizations. We therefore start with $M=1/8$.

\subsection{$M=1/8$}

From Fig. \ref{fig4}(a)-\ref{fig4}(c) we see that the interactions
between the dimers generate a two-dimensional cluster which contains
all dimers in the configuration. In contrast, the dimers in Fig. 
\ref{fig4}(d) form quasi-one-dimensional ``stripe'' clusters.

\begin{figure}[ht!]
\centerline{\epsfig{file=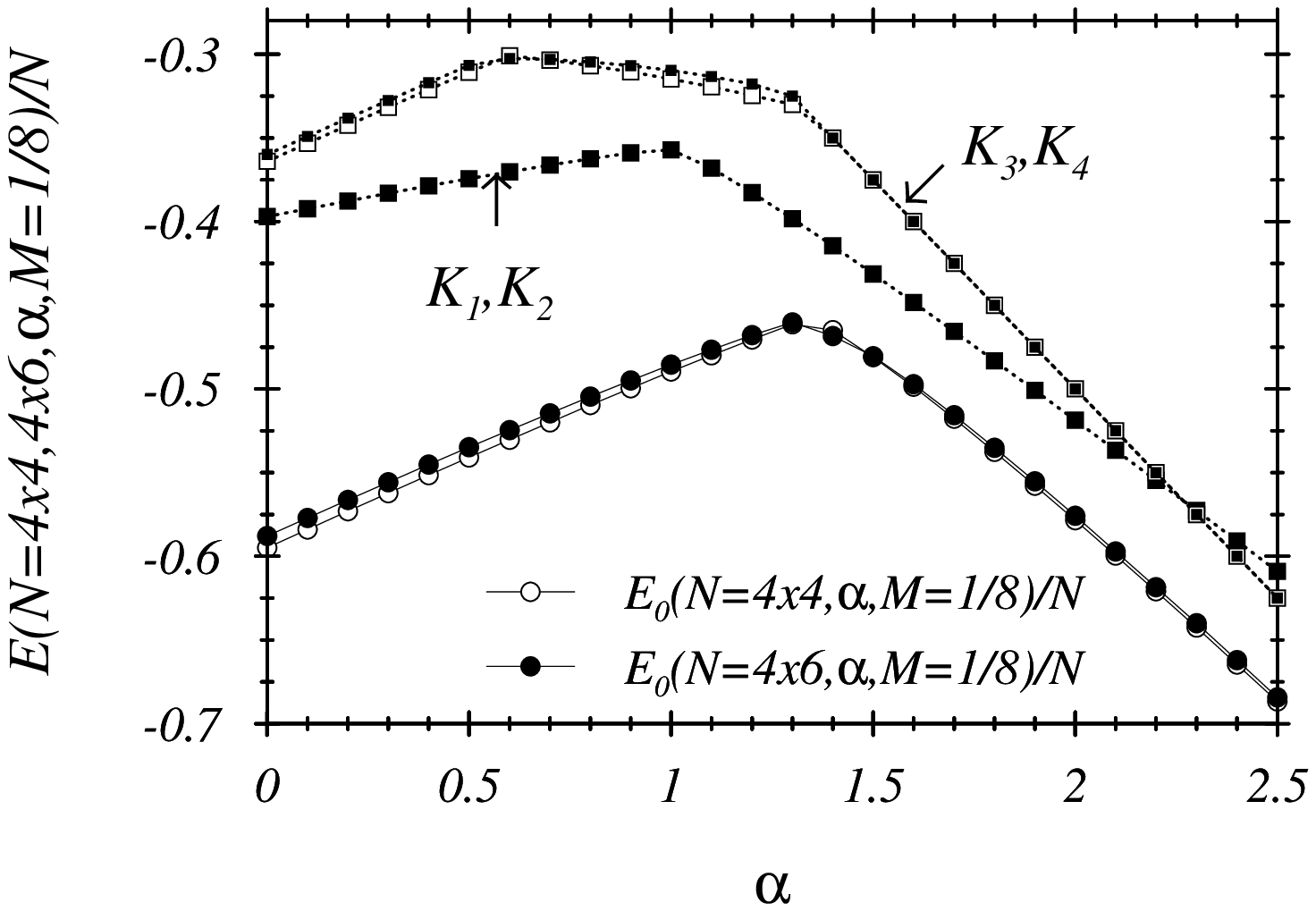,width=8.5cm,angle=0}}
\caption{Ground state energies per site $E_0(N,M,\alpha)/N$ for
Shastry-Sutherland lattices of $N=4\times 4,4\times 6$ sites and
corresponding FMA energies of configurations $K_1,...,K_4$ at
magnetization $M=1/8$.}
\label{fig5x}
\end{figure}

In Fig. \ref{fig5x} the expectation values $E(K_j,M=1/8,\alpha)$, $j=1,2,3,4$
in the frozen monomer approximation -- corresponding to the configurations
Fig. \ref{fig4}(a)-\ref{fig4}(d) -- are represented by dotted lines.

The following points should be noted:

\begin{itemize}

\item
The transition in the magnetic order from configuration $K_2$ [Fig.
\ref{fig4}(b)] to configuration $K_3$ [Fig. \ref{fig4}(c)] occurs
here at a larger value of the frustration parameter
\begin{eqnarray}
\alpha_c(M=1/8) & \simeq & 2.3\,.
\end{eqnarray}
At this value the difference in the expectation values
$E(K_1,M=1/8,\alpha)-E(K_3,M=1/8,\alpha)$
changes sign. For smaller values of $\alpha$ the distribution of
monomers according to Fig. \ref{fig4}(a),(b) is favored in comparison
with the distribution of triplets in Fig. \ref{fig4}(c),(d).

\item
For $\alpha>1.3$ the expectation values
\begin{eqnarray}
E(K_3,M) & = & E(K_4,M)
\end{eqnarray}
corresponding to configurations $K_3$ and $K_4$ [Fig. \ref{fig4}(c)
and Fig. \ref{fig4}(d)] coincide in the frozen monomer
approximation. Indeed, here, the dimer product ansatz 
\eqref{product_ansatz} is an eigenstate of the antiferromagnetic
cluster Hamiltonian \eqref{a4}. In other words: The interactions
between the dimers [dotted lines in Fig. \ref{fig4}(c),(d)] do not
lower the ground state expectation value.

\item
The expectation values $E(K_j,M=1/8)$ deviate significantly for $\alpha
\leq 1.3$ from the exact results given by the solid curves.  Therefore,
other distributions of monomers should play an important role in the
exact ground state.

\item
For small $\alpha$, the eact results show a linear behavior which is well
reproduced in a perturbative expansion in $\alpha$:
\begin{eqnarray}
N^{-1}E_0(M,\alpha) & = & \epsilon_1(M)+\frac{\alpha}{4}\epsilon_2(M)
\end{eqnarray}
where
\begin{eqnarray}
\epsilon_j(M) & = & \langle 0|{\bf S}({\bf x}){\bf S}({\bf y})|0\rangle
\quad j=1,2\\[4pt]
\epsilon_1(1/8) & \simeq & -0.59 \nonumber\\
\epsilon_2(1/8) & \simeq & +0.43 \nonumber
\end{eqnarray}
are the ground state expectation values of the nearest neighbor
($j=1$, $\langle {\bf x},{\bf y}\rangle$) and next-nearest neighbor
($j=2$, $\langle\langle {\bf x},{\bf y}\rangle\rangle$) spin-spin
correlators of the unfrustrated Hamiltonian $H(\alpha=0)=H_1(1,1)$.
\cite{fl01}

\item
Finite-size effects are small, as can be seen from a comparison of
the exact results for the two systems $N=16$ and $N=24$.

\end{itemize}

\subsection{$M=1/6$}

\begin{figure}[ht!]
\centerline{\epsfig{file=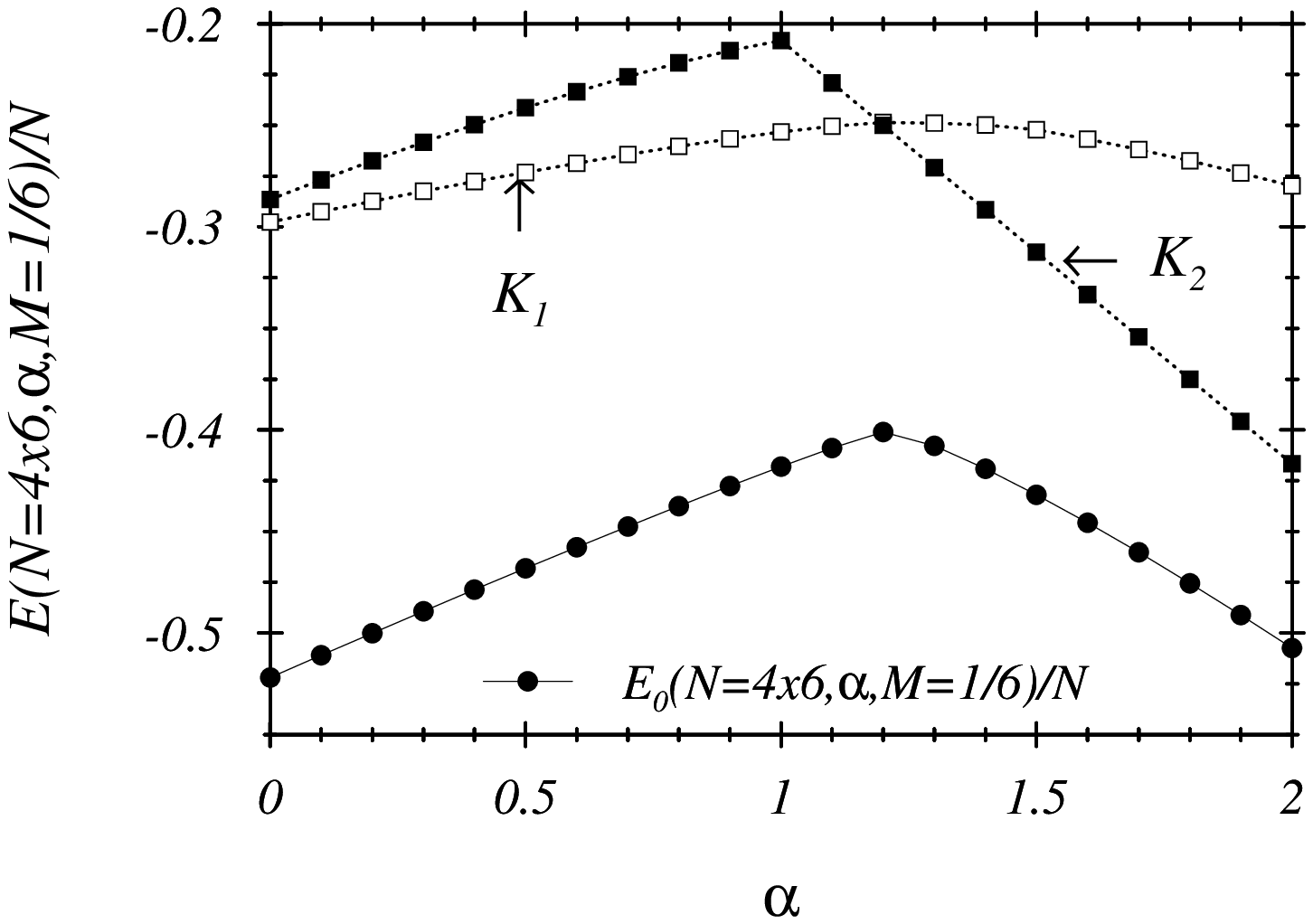,width=8.5cm,angle=0}}
\caption{Ground state energy per site $E_0(N,M,\alpha)/N$ for
a Shastry-Sutherland lattice of $N=4\times 6$ sites and
corresponding FMA energies of configurations $K_1,K_2$ at
magnetization $M=1/6$.}
\label{fig6x}
\end{figure}

In this case the interactions between the dimers form quasi-one-dimensional
clusters with stripe geometry as can be seen from Fig. \ref{fig3}(a),(b).
The expectation values $E(K_j,M=1/6,\alpha)$ $j=1,2$ in the frozen
monomer approximation are shown in Fig. \ref{fig6x}. The transition
point in the magnetic order is found her at
\begin{eqnarray}
\alpha_c(M=1/6) & = & 1.2\,.
\end{eqnarray}

At this point the exact result of the $N=4\times 6=24$ system (solid
line) has its maximum.
The expectation values $E(K_j,M=1/6)$ deviate significantly for $\alpha
< 1.2$ from the exact results given by the solid curve.  


\subsection{$M=1/4$}

The configuration $K_1$ in Fig. \ref{fig2}(a) is built up from 4-point
singlet clusters. In the frozen monomer approximation we lower the
energy if we substitute each dimer pair

\newpage

\begin{figure}[ht!]
\centerline{\epsfig{file=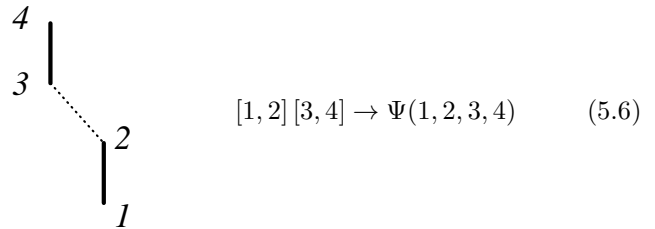,width=4.0cm,angle=0}\hspace{6.0cm}}
\caption{The 2-dimer cluster in the configuration $K_1$
[Fig.\ref{fig2}(a)]}
\label{fig6}
\end{figure}
\vspace{-3.4cm}
\begin{eqnarray} \hspace{2.5cm}[1,2]\,[3,4] & \rightarrow & \Psi(1,2,3,4)
\end{eqnarray}

\vspace{2.4cm}
by the ground state of the 4-point cluster computed from the 4-point
Hamiltonian
\begin{eqnarray}
\label{e_1_4}
H(1,2,3,4) & = & {\bf S}(1){\bf S}(2)+{\bf S}(3){\bf S}(4)+\alpha
{\bf S}(2){\bf S}(3)\,.
\end{eqnarray}
The corresponding ground state energy 
\begin{eqnarray}
E(1,2,3,4) & = & -\frac{2+\alpha}{4}-\frac{1}{2}\left(4-2\alpha+
\alpha^2\right)^{1/2} < -\frac{3}{2}\,
\end{eqnarray}
is lower than the energy of the dimer pair. Taking into account this
effect in the expectation value \eqref{e1m1_4} we find
\begin{eqnarray}
\label{e1m1_4c}
E(K_1,M=1/4) & = & -\frac{N}{16}\left(4-2\alpha+\alpha^2\right)^{1/2}\,.
\end{eqnarray}
Note that there are no interactions between the dimers in the
configuration $K_2$ [Fig. \ref{fig2}(b)]. Therefore the ground state
energy \eqref{e2m1_4} cannot be lowered through the frozen monomer
approximation. The energy differences of \eqref{e1m1_4c} and 
\eqref{e2m1_4} changes its sign at
\begin{eqnarray}
\label{al_c_m1_4} 
\alpha_c(M=1/4) & = & \frac{1}{3}(-1+\sqrt{3})\simeq 0.869\,.
\end{eqnarray}
Below this value the expectation value $E(K_1,M=1/4)$ is a very
poor approximation for the exact ground state energy, indicating
that the true ground state is not reproduced  adequately by the
frozen monomer approximation.

\begin{figure}[ht!]
\centerline{\epsfig{file=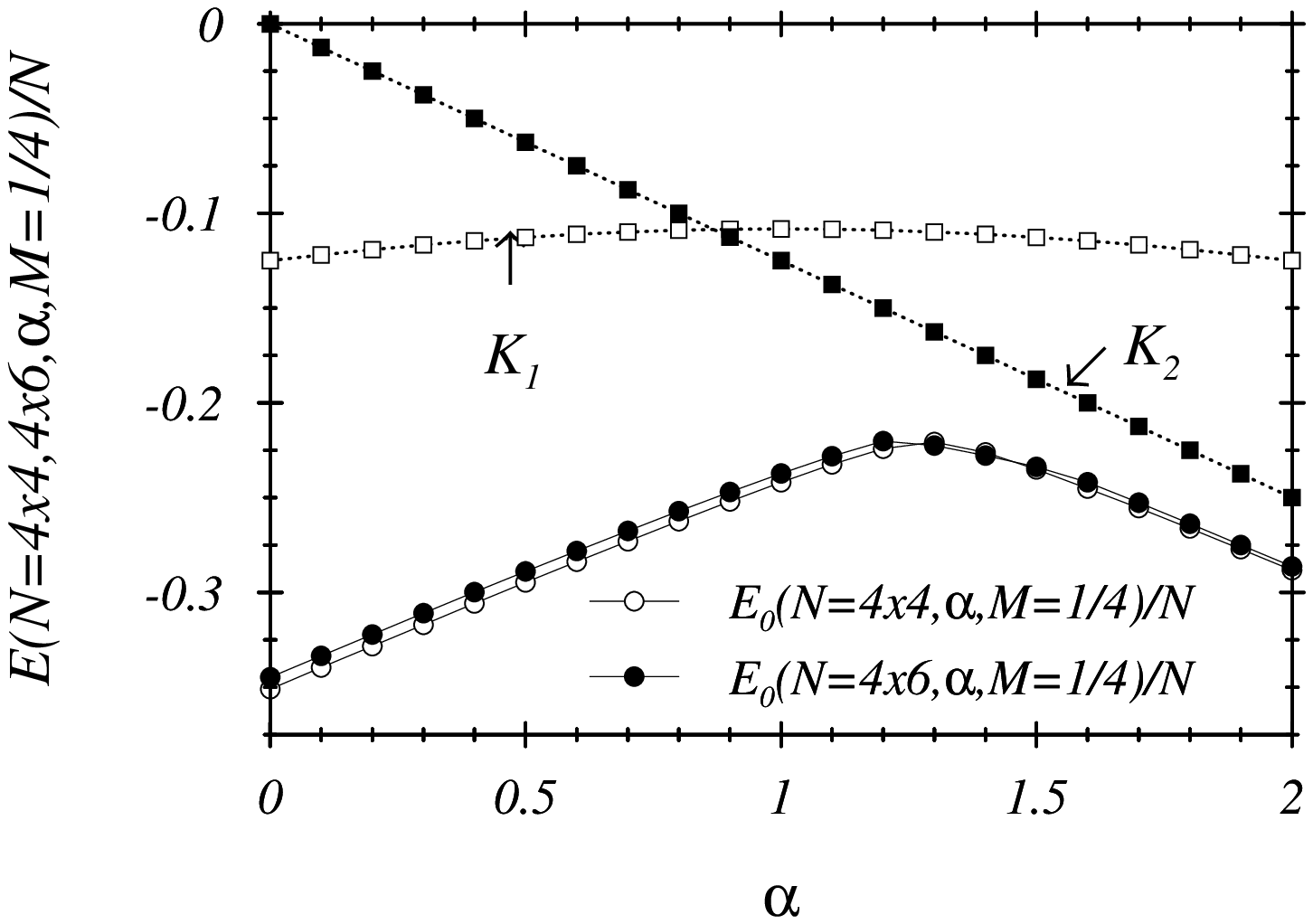,width=8.5cm,angle=0}}
\caption{Ground state energies per site $E_0(N,M,\alpha)/N$ for
Shastry-Sutherland lattices of $N=4\times 4,4\times 6$ sites and
corresponding FMA energies of configurations $K_1,,K_2$ at
magnetization $M=1/4$.}
\label{fig7x}
\end{figure}

In Fig. \ref{fig7x} we compare the energy expectation values
$E(K_1,M=1/4,\alpha)$ \eqref{e1m1_4c} and, $E(K_2,M=1/4,\alpha)$
\eqref{e2m1_4} with the exact diagonalization on finite systems
with $N=4\times 4=16$ and $N=4\times 6=24$ sites. The maximum 
of the exact ground state energy $E_0(M=1/4,\alpha)$
is found at $\alpha\simeq 1.5$ far beyond the transition point
\eqref{al_c_m1_4} from configuration $K_1$ to $K_2$.

\begin{figure}[ht!]
\centerline{\epsfig{file=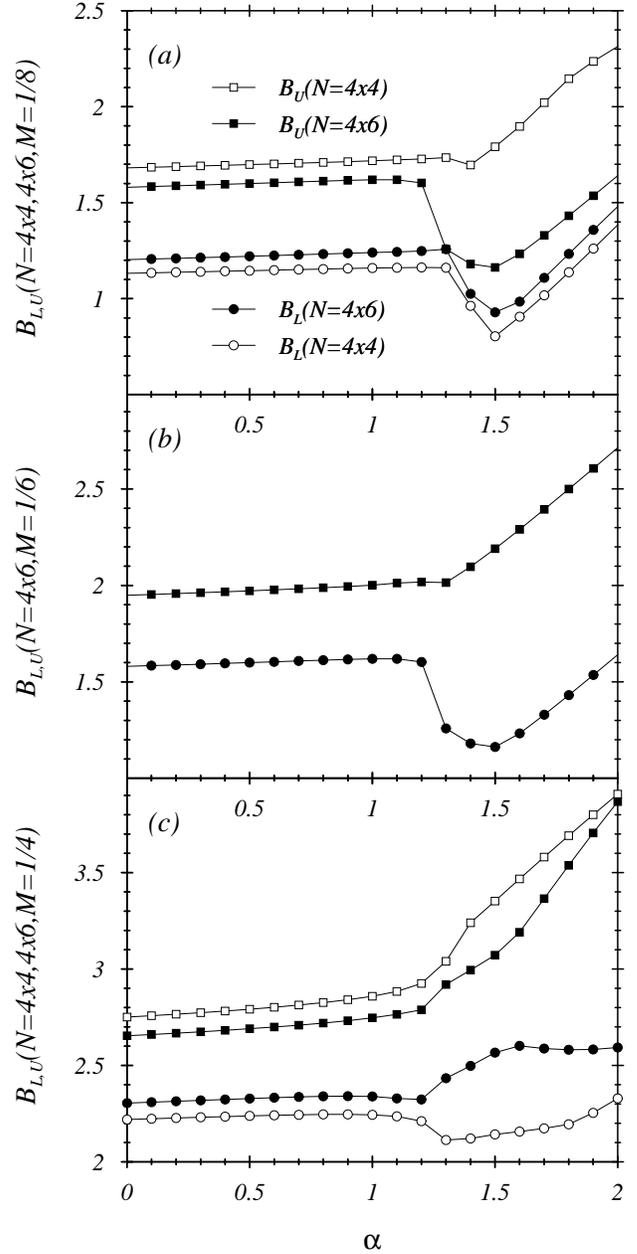,width=8.5cm,angle=0}}
\caption{Upper($B_U$) and lower($B_L$) critical fields for
magnetizations $M=1/8$ (a), $1/6$ (b), $1/4$ (c) calculated on
Shastry-Sutherland lattices of $N=4\times 4$ (a,c) and
$N=4\times 6$ (a-c) sites.}
\label{fig8x}
\end{figure}

We have also studied the formation of plateaus in the magnetization 
curve at $M=1/8,1/6,1/4$ by exact diagonalizations on the finite
clusters with $N=4\times 4=16$ and $N=4\times 6=24$ sites. The
lower and upper critical fields
\begin{eqnarray}
B_L(M,\alpha) & = & E_0(M,\alpha)-E_0(M-2/N,\alpha)\\
B_U(M,\alpha) & = & E_0(M+2/N,\alpha)-E_0(M,\alpha)
\end{eqnarray}
were computed from ground state energies $E_0(M-2/N,\alpha)$, 
$E_0(M,\alpha)$, $E_0(M+2/N,\alpha)$ with neighboring total spins
$S-1,S,S+1$, $S=M\cdot N$. The results are shown in Fig. \ref{fig8x}
(a) for $M=1/8$ (b) for $M=1/6$ (c) for $M=1/4$.

For $\alpha<1.2$ all critical fields are rather $\alpha$-independent.
The finite-size effects indicate that the plateau width
\begin{eqnarray}
\Delta(M,\alpha) & = & B_U(M,\alpha)-B_L(M,\alpha)\quad\mbox{for  }
\alpha<1.2
\end{eqnarray}
will vanish in the thermodynamic limit $N\rightarrow\infty$ as it
is known for the unfrustrated model ($\alpha=0$).

For $\alpha=1.5$ the lower critical fields $B_L(M=1/8,\alpha)$ and
$B_L(M=1/6,\alpha)$ have a pronounced minimum; beyond this value
($\alpha>1.5$) all lower and upper critical fields for $M=1/8,1/6$ 
increase with $\alpha$.

For $M=1/8$ [Fig. \ref{fig8x}(a)] and $\alpha>1.5$, finite-size
effects appear to be small for $B_L(M=1/8,\alpha)$ but large for
$B_U(M=1/8,\alpha)$. 
We suggest that the rectangular geometry of the $4\times 6$ system
might be responsible for this failure. It breaks the rotational
invariance and therefore does not allow for the rotated
patterns in Fig. \ref{fig4}.

For $M=1/4$ [Fig. \ref{fig8x}(c)] and $\alpha>1.5$ we observe
a rather clean signal for the opening of a magnetization plateau.


\section{Discussion and perspectives}

In this paper, we have investigated the magnetic order of the
Shastry-Sutherland model at fixed magnetizations $M=1/8,1/6,1/4$.
For large enough values of the frustration parameter
\begin{eqnarray}
\alpha>\alpha_c(M=1/8)=2.3,& \quad & \alpha>\alpha_c(M=1/6)=1.2,
\nonumber\\
\alpha>\alpha_c(M=1/4)=0.89 & &\nonumber
\end{eqnarray}
configurations built up from singlets and triplets on the
Shastry-Sutherland lattice [cf. Figs. \ref{fig2}(b), \ref{fig3}(b),
\ref{fig4}(c),(d)] -- as they were proposed in
[\onlinecite{kageyama99,onizuka00,momoi00}] -- yield the lowest energy
expectation values. Here, a strong coupling approach
($\alpha^{-1}\rightarrow 0$) to take into account singlet-triplet
interactions is applicable. With this method Momoi and Totsuka
\cite{momoi00} found evidence for plateaus in the magnetization
curve at $M=1/4$ and $M=1/6$.
However, they did not find plateaus at smaller magnetizations
($M=1/8$ and $M=1/16$), ``since the mechanism to stabilize these
plateaus is not yet clear'' -- as they say.

We think that this failure has a simple explanation: The 
singlet-triplet configurations on the Shastry-Sutherland lattice
(cf. e.g. Fig. \ref{fig4}(c),(d) for $M=1/8$) are unfavorable,
since the formation of triplets on next-nearest neighbor sites
costs energy [cf. e.g. \eqref{khk}]. Configurations -- like
$K_2$ in Fig. \ref{fig4}(b) for $M=1/8$ -- with well separated
monomers (unpaired spin-up states) yield a lower energy as long
as $\alpha$ is not too large ($\alpha<\alpha_c(M=1/8)=2.3$).

If the coupling $\alpha$ -- realized in the compound
$SrCu_2(BO_3)_2$ -- is indeed below this value, the experimentally
observed plateaus at $M=1/8$ and $M=1/16$ cannot be associated
with singlet-triplet configurations on the Shastry-Sutherland lattice.

In order to find the correct magnetic order at low magnetizations
$M<\nu/2N$ and $\alpha<2.3$ a more general ansatz for the ground
state configurations is needed. This can be constructed in terms
of monomers at fixed sites ${\bf x}_1\ldots {\bf x}_
{\nu}$. The spins on the remaining sites form an antiferromagnetic
cluster, the ground state energy of which depends on the fixed
positions of the $\nu$ monomers. Therefore, a specific distribution
of monomers characterizing the magnetic order is given by a 
minimum of the ground state energy of the corresponding
antiferromagnetic cluster (cf. e.g. Fig. \ref{fig4}(a),(b) for
$M=1/8$).
We expect that for small values of $M$ -- in particular in the
sectors with a finite number $\nu$ of monomers, i.e. 
$M=\nu/2N\rightarrow 0$ for $N\rightarrow\infty$ -- the singlet-triplet
configurations on the Shastry-Sutherland lattice are dominant
again (for $\alpha\geq\alpha_c(M=0)=1.4$). Each of the $(N-\nu)/2$
singlets lowers the energy by $-3\alpha/4$ whereas each of the
few ($\nu/2$) triplets costs energy $\alpha/4$.

It should also be noted that the frozen monomer approximation
becomes better and better for $M\rightarrow 0$, since the
antiferromagnetic clusters cover more and more of the whole
lattice.

Finally, we have also studied the formation of plateaus in the
magnetization curve of the Shastry-Sutherland model.

We looked for the $\alpha$-dependence of the lower and upper
critical fields as they follow from exact diagonalizations on
finite clusters with $N=4\times 4=16$ and $N=4\times 6=24$
sites. All the critical fields are almost $\alpha$-independent
for $\alpha<1.2$, but change rapidly above this value.
Indications for the opening of a plateau are visible for
$M=1/4,1/6$ supporting previous results with other methods.
\cite{misguich01,momoi00,fukumoto00}

The situation for $M=1/8$ appears to be more subtle. The lower
critical field has a pronounced minimum at $\alpha=1.5$. Here,
the finite-size effects are rather small. In contrast the upper
critical field reveals a strong finite-size dependence.
Computations on larger systems are needed for a reliable estimate
of the thermodynamic limit.


\acknowledgements

We are indebted to M. Karbach for a critical
reading of the manuscript.


 
\end{document}